\documentstyle[12pt,aasms4]{article}

\slugcomment{Astrophysical J. Letters}

\lefthead{Araki et al.}
\righthead{The HCNH$^+$ and its Isotopic Species}

\begin{document}

\title{Laboratory Measurement
of the Pure Rotational Transitions\\
of the HCNH$^+$ and its Isotopic Species}

\author{M. Araki, H. Ozeki and S. Saito}
\affil{The Graduate University for Advanced Studies and Institute for Molecular
 Science,\\ 
Myodaiji, Okazaki 444-8585, Japan}

\begin{abstract}
   The pure rotational transitions of the protonated hydrogen cyanide ion, 
HCNH$^+$, and its isotopic species, HCND$^+$ and DCND$^+$, were measured in 
the 107 - 482 GHz region with a source modulated microwave spectrometer.   The 
ions were generated in the cell with a magnetically confined dc-glow discharge of HCN 
and/or DCN.   The rotational constant $B_0$ and the centrifugal distortion constant 
$D_0$ for each ion were precisely determined by a least-squares fitting to the 
observed spectral lines.   The observed rotational transition frequencies by 
laboratory spectroscopy and the predicted ones are accurate in about 30 to 40 kHz and 
are useful as rest frequencies for astronomical searches of HCNH$^+$ and 
HCND$^+$.
\end{abstract}


\section{INTRODUCTION}

The protonated hydrogen cyanide ion, HCNH$^+$, has drawn much attention because 
of its importance in the chemistry of interstellar clouds.   HCNH$^+$ was first 
detected toward Sgr B2 by Ziurys \& Turner (1986) who observed three rotational 
transitions of $J = 1-0$ to $3-2$.   The fractional abundance of HCNH$^+$ was 
estimated to be $3 \times 10^{-10}$.   Later, Ziurys, Apponi, \& Yoder (1992) tried 
to observe HCNH$^+$ in the dark clods, and succeeded to detect it toward TMC-1 with 
the fractional abundance being estimated to be $3 \times 10^{-9}$.   The hyperfine 
structure of the $J = 1-0$ transition was resolved and the quadrupole coupling 
constant $eQq$ of the nitrogen nucleus was determined.   Interestingly, these 
abundances in the both clouds are generally higher than those predicted by several 
model calculations (\cite{mil97,has93,her89,lan89}).

The main formation reactions of HCNH$^+$ in dark clouds are the following (e.g., 
\cite{sch91});
\begin{eqnarray}
\hbox{H}_3^+ + \hbox{HNC(HCN)} \rightarrow \hbox{HCNH}^+ + 
\hbox{H}_2,\nonumber\\
\hbox{HCO}^+ + \hbox{HNC(HCN)} \rightarrow \hbox{HCNH}^+ + \hbox{CO},\\
\hbox{H}_3\hbox{O}^+ + \hbox{HNC(HCN)} \rightarrow \hbox{HCNH}^+ + 
\hbox{H}_2\hbox{O}\nonumber 
\end{eqnarray}
Furthermore, HCNH$^+$ is a precursor of HCN, HNC, and CN by dissociative 
recombination with electron;
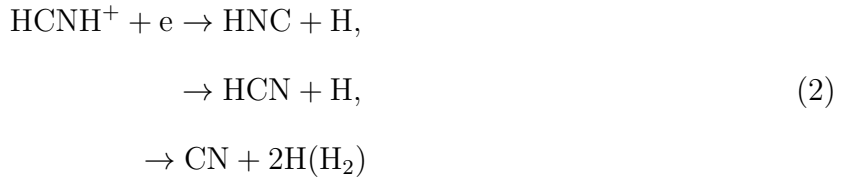
\begin{eqnarray}
\hbox{HCNH}^+ + \hbox{e} \rightarrow	 \hbox{HNC} + \hbox{H},\nonumber \\
	\rightarrow	 \hbox{HCN} + \hbox{H},\\ 
	\rightarrow	 \hbox{CN} + \hbox{2H}(\hbox{H}_2)\nonumber
\end{eqnarray}
HCN, HNC, and CN have been observed abundantly in many interstellar clouds.   
The higher abundance of HNC than that estimated from thermodynamical equilibrium 
in the interstellar cloud is a convincing evidence suggesting the contribution of these 
reactions (\cite{her95,sch92,sch91}).   So far the branching ratio of this reaction was 
predicted from theoretical calculation by Talbi, Ellinger, \& Herbst (1996) and has not 
been measured in laboratory experiment.   It can be said that the HCNH$^+$ ion is 
ubiquitous in molecular clouds and plays an important role in interstellar hydrogen 
cynaide chemistry.

The isotopic species of HCN and HNC are the precursors or the reaction products of 
HCNH$^+$ as seen in reactions (1) and (2), and are also known to show significant 
deuterium enrichment in the interstellar media 
(\cite{how93,wat76,bro77,god77,man91,pen77,tur78}).   The high abundance of the 
deuterium species can be mostly explained by chemical fractionantion at low 
temperature environment.   However the observation of DCN or DNC in the hot 
region like Orion-KL strongly suggests an involvement of dusts in the formation or 
storage process, i.e. sublimation of the fossil molecule from the dusts (\cite{man91}).   
The observation of the deuterium species of the HCNH$^+$ ion, HCND$^+$ and 
DCNH$^+$, is essential to a thorough understanding of the cyanide chemistry in the 
interstellar media.   In that sense, laboratory spectroscopy of HCNH$^+$ and its 
isotopic species is quite important to provide the rest frequencies of the rotational 
transitions.  

There are many laboratory spectroscopic reports for vibration-rotation transitions of 
HCNH$^+$ in the infrared region (e.g., \cite{ama86,kaj88,liu88}).   The $J = 2-1$, 
$3-2$, and $5-4$ transitions of HCNH$^+$ were observed by Bogey et al. (1985) in the 
millimeter and submillimeter wave region and the molecular constants were obtained.   
Based on these results, HCNH$^+$ was detected in interstellar clouds.

On the contrary, the microwave spectroscopic study of the deuterated species has not 
been done.   Bogey, Demuynck, \& Destombes (1985) reported a tentative detection 
of DCNH$^+$ at 370 GHz region ($J = 6-5$), however it has not been confirmed.   
For the vibration-rotation transitions of isotopic species in the infrared region, the 
$\nu_2$ band of HCND$^+$ and the $\nu_1$ band of H$^{13}$CNH$^+$ were 
observed by Amano \& Tanaka (1986) and they determined the $r_s$ structure.   
The $\nu_1$ band of DCNH$^+$ was reported by Amano (1984).   The experimental 
errors of the rotational transitions derived from their molecular constants are about 
several MHz or more, which are not accurate enough to search the molecule in the 
astronomical observation.   This letter reports the laboratory microwave 
spectroscopy of the HCNH$^+$ ion and its deuterated species for providing accurate 
rest frequencies of their rotational transitions.

\section{EXPERIMENTAL DETAILS}

Pure rotational transitions of the HCNH$^+$, HCND$^+$, and DCND$^+$ were 
observed using the same experimental apparatus as that previously described 
(\cite{sai93}).   Briefly, a free space discharge cell with 2 m length and 10 cm outer 
diameter was cooled by circulating liquid nitrogen through a copper tube soldered on a 
copper sheet covering the glass cell.   Around the cell is wound a solenoid of 
enameled wire.   The magnetic field was generated to confine the discharge plasma.   
The spectrometer was a 100 kHz source-modulated system.   Microwave radiation 
was generated with a combination of frequency multipliers and klystrons.   The InSb 
photoconductive detector cooled by liquid helium was used to measure the power of 
microwave radiation.

The HCNH$^+$ was generated by the dc-glow discharge in HCN (1 mTorr) at -100 to -
140$^{\circ}$C.   The DCND$^+$ was with DCN (5 mTorr) and the HCND$^+$ with 
the gas mixture of HCN (2 mTorr) and DCN (5 mTorr).   The method of a 
magnetically confined dc-glow discharge was employed for efficient production of 
protonated ions which was enhanced due to the lengthening of the ion-rich negative 
glow and the increase of ionizing electron density (\cite{del83,bog85}).   The optimum 
magnetic field was about 130 Gauss, and the discharge current and voltage were 40 
mA and 2200 V, respectively.   Signal intensities of the observed lines were enhanced 
due to the magnetic field, confidently indicating the lines to be the transitions of 
protonated ions.   The dependence of the $J = 6-5$ transition of HCND$^+$ on the 
magnetic field is shown in Figure \ref{fig1}.
\placefigure{fig1}

For the HCNH$^+$ ion, three lines were remeasured which were already reported by 
Bogey , Demuynck, \& Destombes (1985) and two other lines were newly measured in 
the present study.   On the basis of their observation toward TMC-1, Ziurys, Apponi, 
\& Yoder (1992) obtained the $eQq$ constant of the nitrogen nucleus to be -0.49 MHz.   
For the observed $J = 2-1$ line in the present study, the hyperfine splitting by the 
$eQq$ interaction of the nitrogen nucleus was not resolved.   The full width at half 
maximum of the $J = 2-1$ transition was about 400 kHz, which is mainly due to 
pressure broadening and not narrow enough to resolve the hyperfine structure.   The 
frequencies of DCND$^+$ were predicted from the $r_s$ structure (\cite{ama86}).   
For HCND$^+$ the frequencies have been predicted from the molecular constants 
determined by the infrared study (\cite{ama86}).   The lines of the HCND$^+$ and 
DCND$^+$ were detected within estimated error ranges.  The frequencies of 
DCNH$^+$ were also predicted from the molecular constants obtained from infrared 
spectra (\cite{ama84}).   Several attempts were made on the detection of DCNH$^+$, 
but without success due to the small dipole moment as discussed later.   All the 
observed line frequencies of three isotopic species are listed in Table \ref{tbl-1}.
\placetable{tbl-1}

The observed frequency may include an ion-drift Doppler shift, and we measured 
difference between the observed frequencies of the $J = 5-4$ transition of 
HCNH$^+$ in normal and opposite electrode configurations, where in the normal one 
microwave radiation propagated from the anode to the cathode, and in the opposite 
one vise versa.   However, corrections due to the drift Doppler shift turned out to be 
unnecessary for all the measured lines because the amount of shift was smaller than 
our experimental accuracy of several tens kilohertz (Table \ref{tbl-1}).

\section{RESULT AND DISCUSSION}

The observed line frequencies of HCNH$^+$, HCND$^+$, and DCND$^+$ were 
analyzed using a rotational energy formula of linear molecule with a least-squares 
method.   The molecular constant for each species was precisely determined.   Two 
transitions of HCNH$^+$ were newly measured in addition to the previous study 
(\cite{bog85}), consequently the precision of the determined rotational and centrifugal 
distortion constants was improved by about one order of magnitude. The standard 
deviation of HCNH$^+$ lines decreased by about one order from the previous values.   
For HCND$^+$, the accuracy of the rotational constant was improved by more than 
two orders of magnitude from the previous infrared study.   The DCND$^+$ ion was 
spectroscopically detected for the first time.   These molecular constants are listed in 
Table \ref{tbl-2}.   The accuracy of the rest transition frequencies is around several 
tens of kilohertz and they are now available for the astronomical search.
\placetable{tbl-2}

In contrast to the above three species the signal of DCNH$^+$ was not detected in this 
laboratory study.   Although we searched the region where Bogey, Demuynck, \& 
Destombes (1985) reported its detection and also that estimated from infrared study, 
the signal intensity was less than our detection limit.   This is due to the small 
electric dipole moment of the molecule.   Botschwina (1986) calculated this physical 
quantity for four isotopic species.   He reported that the dipole moment of 
HCNH$^+$, HCND$^+$, and DCND$^+$ were -0.29, -0.54, and -0.26 D respectively, 
while for DCNH$^+$ it was reduced to 0.004 D.   Our result strongly supports his 
calculation; its dipole moment was estimated to be at least less than 0.1 D if the 
statistical partitioning of the H and D atoms was assumed in the formation reaction in 
the discharge cell.   As a result, the astronomical search for DCNH$^+$ in the 
microwave region seems to be difficult.   The dipole moment of HCND$^+$ is about 
twice larger than that of HCNH$^+$ (\cite{bot86}), so that its search in interstellar 
clouds may be advantageous.

Now we concern about a possibility of the astronomical detection of the deuterated 
species, HCND$^+$.   Deuterium enrichment is observed in many molecules, as 
HCO$^+$, HNC, HCN, and others.   The abundance ratios to their normal species of 
DCO$^+$, DNC, and DCN in TMC-1 are 0.013-0.017 (\cite{gue82}), 0.015 
(\cite{gue82}), and 0.022-0.024 (\cite{woo87}), respectively.   HCNH$^+$ is related 
with HNC and HCN via reactions (1) and (2), and strong deuterium enrichment is 
expected for the ratio of [HCND$^+$]/[HCNH$^+$].   Recently Millar, Bennett, \& 
Herbst (1989) and Howe \& Millar (1993) predicted its ratio by model calculations in 
TMC-1 to be $4.9-8.6 \times 10^{-2}$ and $7.1-10 \times 10^{-2}$ in early time ($3 
\times 10^{5}$ yr) and in steady state, respectively.   We employed the highest value 
in the present calculation, i.e., [HCND$^+$]/[HCNH$^+$] = 0.1.

Radiation temperature $T_R$ of the $J = 1-0$ transition and $\Delta V_{1/2}$ for 
HCNH$^+$ toward TMC-1 were reported by Ziurys, Apponi, \& Yoder (1992): 
$T_R$ = 0.42 K and $\Delta V_{1/2}$ = 0.5 kms$^{-1}$ where the $T_R$ was summed 
over all three hyperfine components.   Column density of HCNH$^+$ was 
recalculated to be $N_{tot} = 5.6 \times 10^{13}$ cm$^{-2}$ using a dipole moment of 
0.29 D (\cite{bot86}) and $T_{rot} = T_{ex} = 7 K$.   Column density of 
HCND$^+$ toward TMC-1 is therefore $5.6 \times 10^{12}$ cm$^{-2}$.   For the $J 
= 2-1$ and $3-2$ transitions, the radiation temperatures in optically thin and local 
thermal equilibrium case are calculated to be about 230 and 150 mK, respectively, 
with the dipole moment of 0.54 D (\cite{bot86}).   The present calculation 
demonstrates a high possibility of the detection of the HCND$^+$ ion toward TMC-1 
as far as the deuterium ratio is around the order of several percent.   The observation 
of HCND$^+$ may help us to understand the HCN/HNC chemistry in the interstellar 
media.

\acknowledgments

The authors thank to Hitoshi Odashima for his comment on the Doppler shift of ion in 
the discharge plasma.

\appendix

\clearpage

\begin{deluxetable}{clrrlrrlr}
\footnotesize
\tablecaption{Observed Rotational Transitions of HCNH$^+$ and Its Isotopic Species 
(MHz) \label{tbl-1}}
\tablehead{
\colhead{Transition} & \multicolumn{2}{c}{HCNH$^+$} & \colhead{} & 
\multicolumn{2}{c}{HCND$^+$} & \colhead{} & \multicolumn{2}{c}{DCND$^+$} \\
\cline{2-3} \cline{5-6} \cline{8-9} \\ 
 \colhead{$J'-J$} &
 \colhead{$\nu_{obs}$\tablenotemark{a}} & 
\colhead{$\Delta\nu$\tablenotemark{b}} & &
 \colhead{$\nu_{obs}$\tablenotemark{a}} & 
\colhead{$\Delta\nu$\tablenotemark{b}} & &
 \colhead{$\nu_{obs}$\tablenotemark{a}} & 
\colhead{$\Delta\nu$\tablenotemark{b}}
}
\startdata
2-1 & 148221.450(17) & $-$0.004 & & 126821.726\tablenotemark{c} & & & 
107095.712(35)& $-$0.010 \nl	
3-2 & 222329.310(6) & 0.020 & & 190230.470(23) & $-$0.012 & & 160642.079(75) & $-
$0.01\phn \nl	
4-3 & 296433.652(17) & $-$0.003 & & 253636.742(34) & 0.032 & & 214186.618(17) & $-
$0.047\tablenotemark{d} \nl	
5-4 & 370533.375(23) & $-$0.020 & & 317039.543(23) & $-$0.025 & & 267728.913(18) & 
0.062\tablenotemark{d} \nl
6-5 & 444627.361(10) & 0.010 & & 380438.217(28) & 0.004 & & 321268.058(10) & 0.008 
\nl	
7-6 & & & & 443831.804(17) & 0.002 & & 374803.689(7) & 0.022 \nl
8-7 & & & & & & & 428335.077(25) & $-$0.024 \nl
9-8 & & & & & & & 481861.762(37) & 0.005 \nl
\enddata

\tablenotetext{a}{Measurement error($\sigma$) is given in the parentheses.}
\tablenotetext{b}{$\Delta\nu = \nu_{obs}- \nu_{calc}$.}
\tablenotetext{c}{Calculated value.   The molecular constants given in 
Table~\ref{tbl-2} are used.}
\tablenotetext{d}{Not included in the least-squares fit.}

\end{deluxetable}

\clearpage

\begin{deluxetable}{clll}
\tablecaption{Molecular Constants of HCNH$^+$ and Its Isotopic Species 
(MHz)\tablenotemark{a} \label{tbl-2}}
\tablehead{
\colhead{Constants} & \colhead{HCNH$^+$} & 
\colhead{HCND$^+$} & \colhead{DCND$^+$} 
}
\startdata
$B_0$ & 37055.7491(74) & 31705.7123(88) & 26774.1295(56) \nl
$D_0$ & \phn \phn \phn \phn 0.048193(136) & \phn \phn \phn \phn 
0.035109(119) & \phn \phn \phn \phn 0.024888(44) \nl
rms (kHz) & \phn \phn \phn 17.8 & \phn \phn \phn 24.6 & \phn \phn \phn 18.5 
\nl
\enddata

\tablenotetext{a}{MHz ($3\sigma$).}

\end{deluxetable}

\clearpage

\figcaption[araki.eps]{ The $J = 6-5$ transition of HCND$^+$ observed in a 
magnetically confined dc-glow discharge where the magnetic field of 145 G in the 
upper spectrum and 0 G in the lower one.   The spectra were obtained by 40 s 
averaging with a time constant of 1 ms. \label{fig1}}

\end{document}